\documentclass[pra,superscriptaddress]{revtex4}
\usepackage{graphicx}
\usepackage{dcolumn}
\usepackage{bm}
\usepackage{amsmath}
\usepackage{amssymb,amsbsy}
\usepackage{color,ulem,cancel}
\newcommand{\beq}{\begin{equation}}
\newcommand{\eeq}{\end{equation}}
\newcommand{\bea}{\begin{eqnarray}}
\newcommand{\eea}{\end{eqnarray}}
\newcommand{\bec}{\begin{center}}
\newcommand{\enc}{\end{center}}
\newcommand{\bfr}{\begin{flushright}}
\newcommand{\efr}{\end{flushright}}
\newcommand{\la}{\langle}
\newcommand{\ra}{\rangle}

\newcommand{\om}{\omega}
\newcommand{\kap}{\kappa}
\newcommand{\gam}{\gamma}

\newcommand{\ve}{\varepsilon}

\newcommand{\tc}{\widetilde{c}}

\newcommand{\bars}{\overline{s}}
\newcommand{\cE}{{\cal E}}

\newcommand{\ha}{\hat{a}} 
\newcommand{\hb}{\hat{b}} 
\newcommand{\hc}{\hat{c}}
\newcommand{\hd}{\hat{d}}
\newcommand{\he}{\hat{e}}
\newcommand{\hs}{\hat{s}}
\newcommand{\hA}{\hat{A}} 
 
\newcommand{\hH}{\hat{H}} 
\newcommand{\hS}{\hat{S}} 
 
\newcommand{\hX}{\hat{X}}
\newcommand{\hY}{\hat{Y}}
\newcommand{\hZ}{\hat{Z}}
\newcommand{\hsig}{\hat{\sigma}}

\newcommand{\hGam}{\hat{\Gamma}}
\newcommand{\geff}{g_\mathrm{eff}}

\begin{document}
\title{Deterministic three-photon down-conversion 
by a passive ultrastrong cavity-QED system}
\author{Kazuki Koshino}
\email{kazuki.koshino@osamember.org}
\affiliation{College of Liberal Arts and Sciences, Tokyo Medical and Dental
University, Ichikawa, Chiba 272-0827, Japan}
\author{Tomohiro Shitara}
\affiliation{College of Liberal Arts and Sciences, Tokyo Medical and Dental
University, Ichikawa, Chiba 272-0827, Japan}
\author{Ziqiao Ao}
\affiliation{Department of Advanced Science and Engineering, 
Waseda University, Tokyo 169-8555, Japan}
\affiliation{Advanced ICT Institute, National Institute of Information and 
Communications Technology, Koganei, Tokyo 184-8795, Japan}
\author{Kouichi Semba}
\altaffiliation[Present address: ]{Institute for Photon Science and Technology, 
The University of Tokyo, Tokyo 113-0033, Japan}
\affiliation{Advanced ICT Institute, National Institute of Information and 
Communications Technology, Koganei, Tokyo 184-8795, Japan}
\date{\today}
\begin{abstract}
In ultra- and deep-strong cavity quantum electrodynamics (QED) systems, 
many intriguing phenomena that do not conserve the excitation number 
are expected to occur.
In this study, we theoretically analyze the optical response 
of an ultrastrong cavity-QED system in which an atom is coupled to 
the fundamental and third harmonic modes of a cavity, 
and report the possibility of deterministic three-photon down-conversion 
of itinerant photons upon reflection at the cavity. 
In the conventional parametric down-conversion, a strong input field is needed
because of the smallness of the transition matrix elements 
of the higher order processes. 
However, if we use an atom-cavity system in an unprecedentedly strong-coupling region, 
even a weak field in the linear-response regime is sufficient to cause 
this rare event involving the fourth order transitions.
\end{abstract}
\maketitle

\section{introduction} 
The history of cavity quantum electrodynamics (QED) 
parallels with the enhancement of the atom-cavity coupling $g$. 
From the observations of the suppressed or enhanced atomic decay rate
in the weak coupling regime ($g<\kap, \gam$, 
where $\kap$ and $\gam$ are the loss rates of 
the cavity and the atom, respectively)~\cite{Pur1,Pur2,Pur3,Pur4,Pur5}, 
more than a decade was required to reach 
the strong coupling regime ($g>\kap, \gam$), 
where the Rabi splitting or oscillation becomes observable~\cite{sc1,sc2,sc3,sc4,sc5,sc6}. 
Recently, the ultrastrong coupling regime 
($g\gtrsim \om_a/10, \om_c/10$, where $\om_a$ and $\om_c$ 
are the resonance frequencies of the atom and cavity, respectively) 
and even the deep-strong coupling regime ($g\gtrsim \om_a, \om_c$) 
have been realized in various physical platforms such as 
polaritons, superconducting qubits, and molecules~\cite{usc1,usc2,usc3,usc4,usc5,usc6,usc7}.
In the ultra- and deep-strong coupling regimes, 
various novel phenomena originating in 
the counter-rotating terms of the atom-cavity coupling 
are expected to become observable.

One of such phenomena is the deterministic nonlinear optics~\cite{dno1,dno2,dno3}. 
The generation efficiency of entangled photons, 
which is typically of the order of $10^{-6}$ 
for spontaneous parametric down-conversion
with a bulk nonlinear crystal~\cite{spdc1,spdc2,spdc3}, 
might be drastically improved by this scheme.  
However, the proposed deterministic nonlinear-optical processes are for intracavity photons. 
In order to apply this scheme for itinerant photons, 
deterministic capturing of propagating photons into a cavity is indispensable. 
This is in principle possible 
but requires a precise dynamic control of the external cavity loss rate 
in accordance with the incoming photon shape~\cite{cr1,cr2}. 
In contrast, deterministic down-conversion of itinerant photons is possible 
and has been demonstrated in a passive waveguide QED setup~\cite{wqed1,wqed2,wqed3}. 
The physical origin of the deterministic conversion is 
the destructive interference between incoming field and radiation from the atom. 
Accordingly, the efficiency is sensitive to 
the external loss rates of the relevant transitions of the atom.

\begin{figure}[b]
\begin{center}
\includegraphics[width=70mm]{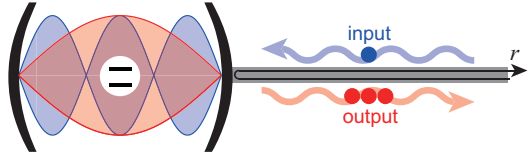}
\end{center}
\caption{
Schematic of the considered setup.
A qubit (resonance frequency $\om_q$) interacts with 
the first ($\om_1$) and third ($\om_3$) cavity modes
with coupling constants $g_1$ and $g_3$, respectively. 
Dissipation channels are as follows: 
qubit radiative decay (rate $\gam$), 
external and internal losses of the first cavity mode ($\kap_{1e}$, $\kap_{1i}$)
and those for the third one ($\kap_{3e}$, $\kap_{3i}$).
A weak monochromatic field (frequency $\om_\mathrm{in}\sim \om_3$) 
is input through the waveguide.
}
\label{fig:setup}
\end{figure}

In this study, we investigate an ultrastrong cavity QED system
in which an atom is placed at the center of the cavity 
and is thus coupled to the fundamental 
and third-harmonic modes of the cavity (Fig.~\ref{fig:setup})~\cite{BSS1}.
We show that, if the atom-cavity and cavity-waveguide couplings are adequately chosen,  
an input photon (resonant to the third mode) is down-converted nearly deterministically 
to three daughter photons (resonant to the fundamental mode) upon reflection at the cavity .
The drastic enhancement of the conversion efficiency 
in comparison with the prior demonstrations of triplet-photon 
generation~\cite{3dc_a1,3dc_a2,3dc_a3,3dc_a4,3dc_b1,3dc_b2,3dc_c1,3dc_c2}
originates in the waveguide QED effect, 
in other words, the engineered dissipation rates of the optical system. 
Such deterministic conversion is possible 
even in the conventional strong-coupling cavity QED. 
However, considering the required intrinsic loss rates of the cavity modes, 
this phenomenon is characteristic to the ultrastrong cavity QED.

The rest of this paper is organized as follows.
We present the theoretical model of the atom-cavity-waveguide 
coupled system in Sec.~\ref{sec:2}. 
We observe the internal dynamics of the atom-cavity system 
and evaluate the effective coupling between the two levels 
relevant to three-photon down-conversion in Sec.~\ref{sec:3}. 
We develop the input-output formalism applicable to 
the ultrastrong coupling regime in Sec.~\ref{sec:4}, 
and apply to the investigated setup in Sec.~\ref{sec:5}.
We numerically show that the deterministic three-photon 
down-conversion is possible in Sec.~\ref{sec:6}
and clarify the required conditions.
Section~\ref{sec:7} is devoted to a summary.

\section{system}
\label{sec:2}
In this study, we investigate an optical response of 
a cavity QED system schematically illustrated in Fig.~\ref{fig:setup}.
A two-level system (qubit) is placed at the center of a cavity
and interacts with the first- and third-harmonic cavity modes.
These cavity modes are coupled to an external waveguide field 
through the right mirror (capacitor, in circuit QED implementation), 
and a monochromatic field close to the resonance of 
the third cavity mode is applied through the waveguide.

The Hamiltonian of the qubit-cavity system is given, setting $\hbar=1$, by
\bea
\hH_s &=& \om_q \hsig^{\dag}\hsig + \om_1 \ha_1^{\dag}\ha_1 + \om_3 \ha_3^{\dag}\ha_3
+ g_1\hX\hZ + g_3\hY\hZ,
\eea
where $\hsig$, $\ha_1$ and $\ha_3$ respectively denote
the annihilation operators of the qubit, 
the first and third cavity modes, 
and $\hX = \ha_1^{\dag}+\ha_1$, $\hY = \ha_3^{\dag}+\ha_3$
and $\hZ = \hsig^{\dag}+\hsig$.
$\om_q$, $\om_1$ and $\om_3$ denote their bare resonance frequencies, 
and $g_1$ and $g_3$ denote the qubit-cavity coupling strengths. 
Note that the counter rotating terms are retained
in order to treat the ultrastrong coupling regime.

We consider five dissipation channels of this system:
the external/internal decay of the first/third cavity mode and 
the longitudinal decay of the qubit. 
We label these dissipation channels as $1e$, $1i$, $3e$, $3i$ and $q$, respectively,
and denote their rates as $\kap_{1e}$, $\kap_{1i}$, $\kap_{3e}$, 
$\kap_{3i}$, and $\gam$, respectively. 
As we observe later (Fig.~\ref{fig:loss}), 
the investigated phenomenon is robust against the qubit dissipation, 
since the qubit excited state is used only virtually. 
We therefore neglect the qubit pure dephasing for simplicity, 
which plays essentially the same role as the longitudinal decay in the present phenomenon.  
The Hamiltonians describing the $1e$ and $3e$ channels are given by
\bea
\hH_{1e} &=& 
\int_0^{\infty} dk \left[
\om_k\hc_k^{\dag}\hc_k + \xi^{(1e)}_k \hX(\hc_k^{\dag}+\hc_k)
\right], \label{eq:H1e}
\\
\hH_{3e} &=& 
\int_0^{\infty} dk \left[
\om_k\hd_k^{\dag}\hd_k + \xi^{(3e)}_k \hY(\hd_k^{\dag}+\hd_k)
\right], \label{eq:H3e}
\eea
where $\hc_k$ ($\hd_k$) is the annihilation operator of 
a waveguide mode with wavenumber $k$ coupled to the first (third) cavity mode
and $\om_k$ is its frequency. 
The commutators for $\hc_k$ and $\hd_k$ are given by
$[\hc_k, \hc^{\dag}_{k'}]=[\hd_k, \hd^{\dag}_{k'}]=\delta(k-k')$.
Although these modes can be treated as a single continuum in principle,  
we may safely treat them as independent continua
because of large separation between the relevant frequencies.
The dispersion relation in the waveguide is linear, $\om_k=k$,  
where the velocity of waveguide photons is set to unity for simplicity.
A real quantity $\xi^{(je)}_k$ ($j=1$, $3$) represents the cavity-waveguide coupling. 
By naively applying the Fermi golden rule, 
the coupling and the decay rate are related by
\bea
\kap_{je} &=& 2\pi \xi_{\om_j}^{(je)2}.
\label{eq:kapje}
\eea
The other loss channels are modeled similarly.
For example, the longitudinal relaxation of the qubit is modeled by 
\bea
\hH_{q} &=& 
\int_0^{\infty} dk \left[
\om_k\he_k^{\dag}\he_k + \xi^{(q)}_k \hZ(\he_k^{\dag}+\he_k)
\right]. \label{eq:Hq}
\eea
The relation to the qubit decay rate is 
$\gam = 2\pi \xi_{\om_q}^{(q)2}$.

\section{coupling between $|g01\ra$ and $|g30\ra$}
\label{sec:3}
\begin{figure}[t]
\begin{center}
\includegraphics[width=80mm]{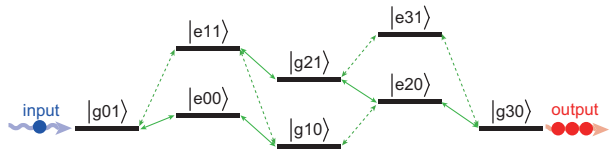}
\end{center}
\caption{
Transition paths between $|g01\ra$ and $|g30\ra$. 
Solid (dotted) arrows indicate the transitions 
conserving (non-conserving) the excitation number. 
}
\label{fig:tran}
\end{figure}
\begin{figure}
\begin{center}
\includegraphics[width=140mm]{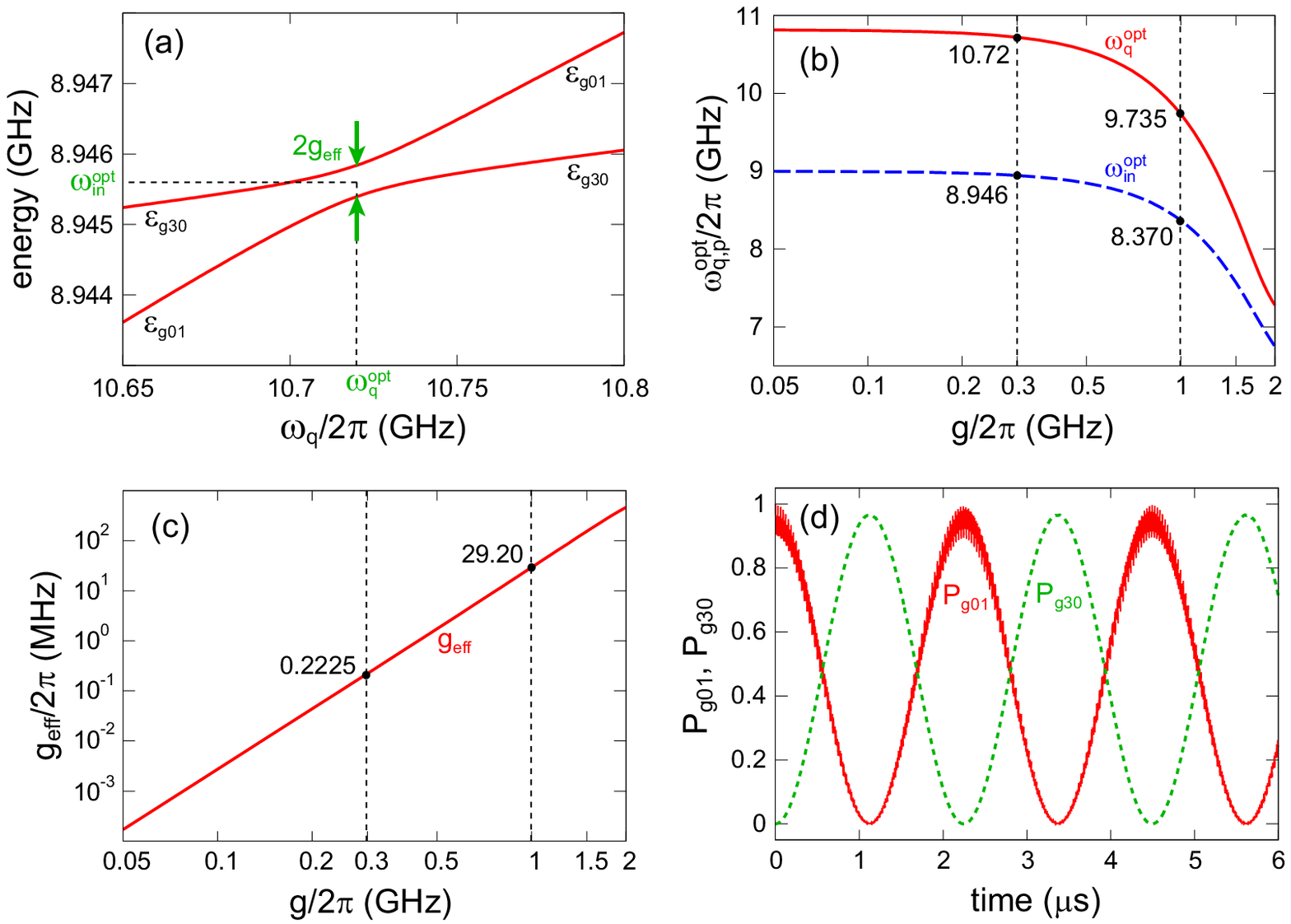}
\end{center}
\caption{
Effective coupling between $|g01\ra$ and $|g30\ra$. 
$3\om_1=\om_3=2\pi\times 9$~GHz and $g_1=g_3(=g)$ are assumed.
(a)~Anticrossing between $|g01\ra$ and $|g30\ra$ varying the qubit frequency $\om_q$. 
$g=2\pi\times 0.3$~GHz.
(b, c)~Dependences of ~$\om_q^\mathrm{opt}$, $\om_\mathrm{in}^\mathrm{opt}$, 
and $\geff$ on the qubit-cavity coupling $g$. 
Specific values at $g=2\pi\times 0.3$ and $1$~GHz are indicated. 
(d)~Vacuum Rabi oscillation between $|g01\ra$ and $|g30\ra$,
starting from $|g01\ra$. 
$g=2\pi\times 0.3$~GHz and $\om_q=\om_q^\mathrm{opt}=2\pi\times 10.72$~GHz.
}
\label{fig:anti}
\end{figure}
In this section, we investigate the properties of the qubit-cavity system, 
neglecting dissipation for the moment. 
We denote the state vector of the system by $|qmn\ra$, 
where $q(=g,e)$ specifies the qubit state
and $m, n$ $(=0, 1, \cdots)$ specify the photon numbers 
of the first and third cavity modes, respectively.
In particular, we focus on the coupling between $|g01\ra$ and $|g30\ra$, 
which is essential for the three-photon down conversion.  
Figure~\ref{fig:tran} shows the transition paths between $|g01\ra$ and $|g30\ra$. 
We observe that $|g01\ra$ and $|g30\ra$ are coupled through 
the fourth order process in the qubit-cavity coupling $g$, 
and that transitions non-conserving the excitation number 
(dotted arrows in Fig.~\ref{fig:tran}) are indispensable for their coupling.

For a strong coupling between these two states, 
degeneracy of these two states is required. 
The eigenenergies of $|g01\ra$ and $|g30\ra$ measured from $|g00\ra$, 
which we denote by $\ve_{g01}$ and $\ve_{g30}$, 
are renormalized by the dispersive qubit-cavity coupling
and depend on $g_{1,3}$ and $\om_q$. 
In Fig.~\ref{fig:anti}(a), by numerically diagonalizing $\hH_s$
(taking into account up to the 6 (2) photon states for $\ha_1$ ($\ha_3$) mode),
$\ve_{g01}$ and $\ve_{g30}$ are plotted as functions of the qubit frequency $\om_q$. 
We observe an anticrossing between them, 
and from this plot we can identify the following quantities. 
(i)~The optimal qubit frequency $\om_q^\mathrm{opt}$, 
at which the level spacing is minimized.
(ii)~The effective coupling $\geff$ between $|g01\ra$ and $|g30\ra$, 
which is half of the level spacing at $\om_q^\mathrm{opt}$.
(iii)~The optimal input photon frequency $\om_\mathrm{in}^\mathrm{opt}$, 
which is the average of the two transition 
frequencies at $\om_q^\mathrm{opt}$.
In Figs.~\ref{fig:anti}(b) and (c), 
assuming $3\om_1=\om_3=2\pi\times 9$~GHz and $g_1=g_3(=g)$, 
we plot $\om_q^\mathrm{opt}$, $\om_\mathrm{in}^\mathrm{opt}$ and $\geff$
as functions of $g$.

We can confirm in Fig.~\ref{fig:anti}(c) that
$\geff$ is proportional to $g^4$, 
which is consistent with the fact that $|g01\ra$ and $|g30\ra$ 
are coupled through the fourth order process. 
We also observe in Fig.~\ref{fig:anti}(b) that,
in the weak-coupling limit of $g \to 0$, 
$\om_\mathrm{in}^\mathrm{opt} \to \om_3=3\om_1=2\pi\times 9$~GHz 
and $\om_q^\mathrm{opt} \to \sqrt{(3\om_3^2-\om_1^2)/2}=2\pi\times 10.82$~GHz.
This optimal qubit frequency in the weak-coupling limit 
is derived as follows.
Within the second-order perturbation in $g_1$ and $g_3$, 
$\ve_{g01}$ and $\ve_{g30}$ are renormalized as
$\ve_{g01} = \om_3-\frac{g_3^2}{\om_q-\om_3}-\frac{g_3^2}{\om_q+\om_3}$ and 
$\ve_{g30} = 3\left( \om_1-\frac{g_1^2}{\om_q-\om_1}
-\frac{g_1^2}{\om_q+\om_1} \right)$, respectively.
The degeneracy condition, $\ve_{g01}=\ve_{g30}$, 
reduces to $\om_q = \sqrt{(3\om_3^2-\om_1^2)/2}$ 
for the present case of $\om_3=3\om_1$ and $g_1=g_3$.

In Fig.~\ref{fig:anti}(d), coherent time evolution of the system 
starting from $|g01\ra$ is shown. 
We observe the vacuum Rabi oscillation between $|g01\ra$ and $|g30\ra$. 
The oscillation is not pure sinusoidal, since 
other states than $|g01\ra$ and $|g30\ra$ (such as $|e00\ra$) 
are involved in forming the eigenstates. 
The Rabi oscillation period measured in 
Fig.~\ref{fig:anti}(d) is $T=2.247$~$\mu$s. 
This is compatible with $\geff=2\pi\times 222.5$~kHz 
evaluated from Fig.~\ref{fig:anti}(a), because $\geff T/\pi \approx 1$.

\section{input-output formalism for ultrastrong cavity QED}
\label{sec:4}
In the following part of this paper, 
we analyze the response of the qubit-cavity system to a waveguide field. 
For this purpose, the input-output formalism is useful. 
However, although highly useful in the weak- and (usual) 
strong-coupling regimes of cavity QED, 
the conventional formalism based on the white reservoir approximation
has several difficulties in treating the ultrastrong cavity QED~\cite{inout_rev1}.
The input-output formalism applicable to the ultrastrong cavity QED
has been discussed first in linear systems~\cite{inout_lin}
and later extended to nonlinear systems~\cite{inout_NL1,inout_NL2}.
However, assuming weak dissipation, 
the counter-rotating terms in the system-environment coupling 
have been neglected in prior works. 
In this section, in order to extend the formalism 
applicable to highly dissipative cases,
we develop an input-output formalism  
incorporating the counter-rotating terms in the system-environment coupling. 

\subsection{Heisenberg equations}
In this section, for simplicity, 
we consider a case in which the cavity QED system is coupled only to 
the $1e$ dissipation channel. 
Namely, the overall Hamiltonian is given by $\hH = \hH_s + \hH_{1e}$. 
We also omit the superscript $(1e)$ throughout this section
(for example, $\xi_k^{(1e)} \to \xi_k$).

The Heisenberg equation of $\hc_k$ is given by
$d\hc_k/dt = i[\hH, \hc_k] = -ik\hc_k -i\xi_k\hX$,
which is formally integrated as
\bea
\hc_k(t) &=& \hc_k(0)e^{-ikt}-i\xi_k \int_0^t d\tau \ e^{-ik\tau} \hX(t-\tau).
\label{eq:ckt}
\eea
The Heisenberg equation of an arbitrary system operator $\hS$ is given by
$d\hS/dt = i[\hH_s, \hS]+i\int_0^{\infty}dk \ \xi_k [\hX,\hS](\hc_k^{\dag}+\hc_k)$. 
Using Eq.~(\ref{eq:ckt}), this becomes a delay differential equation,
\bea
\frac{d}{dt}\hS &=&
i[\hH_s, \hS] + \int_0^{\infty} \!\! dk \int_0^t \!\! d\tau \ \xi_k^2
\left(
e^{-ik\tau}[\hX,\hS]\hX(t-\tau)
-e^{ik\tau}\hX(t-\tau)[\hX,\hS]
\right)
+i[\hX,\hS]\hGam(t) + i\hGam^{\dag}(t)[\hX,\hS],
\label{eq:dSdt1}
\eea
where $\hGam(t)$ is a noise operator defined by
\bea
\hGam(t) &=& \int_0^{\infty} dk \ \xi_k e^{-ikt} \hc_k(0).
\eea
In Eq.~(\ref{eq:dSdt1}) and hereafter, 
we omit explicit time-dependence for the operators 
at time $t$ in the Heisenberg picture. 
Note that the noise operator $\hGam(t)$ is 
{\it not} in the Heisenberg picture.

In order to convert a delay differential equation~(\ref{eq:dSdt1}) 
into an instantaneous one, 
we employ a {\it free-evolution} approximation
for the time evolution of the system operator 
during the delay time $\tau$~\cite{fea0,fea1,fea2}. 
We denote the eigenstates and eigenenergies of $\hH_s$
by $|j\ra$ and $\ve_j$ ($j=0, 1, \cdots$) in the energy-increasing order, 
and define the transition operator by $\hs_{ij}=|i\ra \la j|$.
We then approximate $\hX(t-\tau)$ as
\bea
\hX(t-\tau) &=& \sum_{i,j} x_{ij}\hs_{ij}(t-\tau)
\approx \sum_{i,j} x_{ij} e^{i\ve_{ji}\tau} \hs_{ij}(t),
\label{eq:fev}
\eea
where $x_{ij}=\la i|\hX|j\ra$ and $\ve_{ji}=\ve_j-\ve_i$.
Then, Eq.~(\ref{eq:dSdt1}) is rewritten as
\bea
\frac{d}{dt}\hS &=&
i[\hH_s, \hS] + [\hX,\hS]\hA - \hA^{\dag}[\hX,\hS]
+i[\hX,\hS]\hGam(t) + i\hGam^{\dag}(t)[\hX,\hS],
\label{eq:dSdt2}
\eea
where
\bea
\hA &=& \sum_{i,j}x_{ij}h_{ji}\hs_{ij},
\eea
and $h_{ji} = \int_0^t \! d\tau \int_0^{\infty} \! dk \ \xi_k^2 e^{i(\ve_{ji}-k)\tau}$.
Strictly speaking, this quantity depends on $t$. 
However, since 
$\int_0^{\infty} \!\! dk \ \xi_k^2 e^{-ik\tau}$ is nonzero
only for short $\tau$ ($\sim$ inversed bandwidth of $\xi_k^2$), 
we can safely treat $h_{ji}$ as a $t$-independent quantity,
$h_{ji} \approx \int_0^{\infty} \! d\tau \int_0^{\infty} \! dk \ \xi_k^2 e^{i(\ve_{ji}-k)\tau}$.
Then we have
\bea
h_{ji} &=& -i\int_0^{\infty}dk \frac{\xi_k^2}{k-\ve_{ji}-i0}
=\pi \theta(\ve_{ji})\xi^2_{\ve_{ji}}
-i\mathrm{P}\int_0^{\infty}dk \frac{\xi_k^2}{k-\ve_{ji}},
\label{eq:hji}
\eea
where $\theta$ is the Heaviside step function and 
$\mathrm P$ represents the principal value. 
This is a self energy correction to the transition frequency: 
the real part corresponds to half of the decay rate and 
the imaginary part corresponds the Lamb shift~\cite{Mahan}.
Note that the real part vanishes for negative $\ve_{ji}$, 
reflecting the prohibited decay from a lower level to an upper level.

\subsection{Waveguide-field operator}
For a semi-infinite waveguide (Fig.~\ref{fig:setup}), 
the waveguide eigenmodes are the standing wave extending in the $r>0$ region. 
Assuming an open boundary condition at $r=0$, 
the eigenmode function $f_k$ with wavenumber $k$ is given by
$f_k(r)=\sqrt{2/\pi}\cos(kr)\theta(r)$, 
which is orthonormalized as $\int_0^{\infty}dr f_k(r)f_{k'}(r)=\delta(k-k')$.
Accordingly, the waveguide mode operator $\hc_r$ 
in the real-space representation is defined in the $r>0$ region by 
$\hc_r(t)=\sqrt{2/\pi}\int_0^{\infty} dk \cos(kr) \hc_k(t)$, 
which satisfies the commutation relation $[\hc_r,\hc_{r'}^{\dag}]=\delta(r-r')$.
The incoming field propagates in the $r>0$ region into the negative direction ($k<0$).

However, it is convenient to treat the incoming field
as if it propagates in the $r<0$ region into the positive direction ($k>0$). 
We therefore introduce the real-space representation by
\bea
\tc_r(t) &=& \frac{1}{\sqrt{2\pi}} \int_0^{\infty} dk \ e^{ikr} \hc_k(t),
\eea
where $r$ runs over the full one-dimensional space ($-\infty<r<\infty$)~\cite{fn3}. 
From Eq.~(\ref{eq:ckt}), we have
\bea
\tc_r(t) &=& \tc_{r-t}(0) -\frac{i}{\sqrt{2\pi}} 
\int_0^{\infty} \!\! dk \int_0^t \!\! d\tau \ \xi_k e^{ik(r-\tau)} \hX(t-\tau).
\label{eq:tcr2}
\eea
This equation represents the waveguide-field operator 
in terms of the input-field operator and the system operator, 
and enables, for example, 
evaluation of the output field amplitude and flux.

\subsection{Input-output relation}
We can further simplify Eq.~(\ref{eq:tcr2}) under some approximations. 
Introducing $\tau'=\tau-r$ and 
employing the free-evolution approximation, 
$\hX(t-r-\tau')=\sum_{i,j}x_{ij}\hs_{ij}(t-r) e^{i\om_{ij}\tau'}$, 
Eq.~(\ref{eq:tcr2}) is rewritten as
\bea
\tc_r(t) &=& \tc_{r-t}(0) + \sum_{i,j} x_{ij} f(\ve_{ji},r,t) \hs_{ij}(t-r),
\label{eq:tcr}
\\
f(\ve,r,t) &=& \frac{1}{\sqrt{2\pi}} \int_0^{\infty} dk \frac{\xi_k}{k-\ve}
\left[ e^{i(k-\ve)(r-t)} - e^{i(k-\ve)r} \right].
\label{eq:fr1}
\eea
Note that the integrand in the right-hand side of Eq.~(\ref{eq:fr1})
is not singular at $k=\ve$. 
We can approximately evaluate $f(\ve,r,t)$ as follows.
Since the main contribution of this integral 
comes from the $k \approx \ve$ region, 
we set $\xi_k \approx \xi_{\ve}\theta(\ve)$
and remove the lower limit of $k$ integral as 
$\int_0^{\infty} \approx \int_{-\infty}^{\infty}$. 
Then we have
\bea
f_\mathrm{app}(\ve,r,t) &=& 
-i\sqrt{2\pi}\xi_{\ve} \theta(\ve)\theta(r)\theta(t-r).
\label{eq:fr2}
\eea
In Appendix~\ref{app:comp}, we observe fairly good agreement 
between $f(\ve,r,t)$ and $f_\mathrm{app}(\ve,r,t)$,
assuming a concrete form of $\xi_k$ [Eq.~(\ref{eq:xik1e})]. 
Thus we have
\bea
\tc_r(t) &=& \tc_{r-t}(0) -i\sqrt{2\pi}\theta(r)\theta(t-r)  
\sum_{i<j} \xi_{\ve_{ji}} x_{ij} \hs_{ij}(t-r).
\label{eq:tcr3}
\eea
Note that the summation over $i$ and $j$ is conditioned by $i<j$ in Eq.~(\ref{eq:tcr3}), 
which is due to the following reasons.
(i)~For $i>j$, $\ve_{ji}$ is negative and accordingly $\theta(\ve_{ji})=0$.  
(ii)~For $i=j$, $x_{ij}=0$ due to the parity selection rule.
%
%
Defining the input and output field operators by
$\hc_\mathrm{in}(t)=\tc_{-0}(t)=\tc_{-t}(0)$ 
and $\hc_\mathrm{out}(t)=\tc_{+0}(t)$, 
Eq.~(\ref{eq:tcr3}) is rewritten into a more familiar form,
\bea
\hc_\mathrm{out}(t) &=& \hc_\mathrm{in}(t) -i\sqrt{2\pi} 
\sum_{i<j} \xi_{\ve_{ji}} x_{ij} \hs_{ij}(t).
\label{eq:cout}
\eea

\section{optical response theory}
\label{sec:5}
\subsection{Initial state vector} 
In this study, instead of a single photon pulse, 
we apply a weak classical monochromatic field 
close to the resonance of the third cavity mode
[$\cE_\mathrm{in}(t)=E_\mathrm{in} e^{-i\om_\mathrm{in} t}$ with $\om_\mathrm{in} \sim \om_3$]
to the cavity (Fig.~\ref{fig:setup}). 
Assuming that, at the initial moment ($t=0$),
the overall system is in the vacuum state 
except the applied field, the initial state vector is written as
\bea
|\psi_i\ra &=& \exp\left(
\sqrt{2\pi}E_\mathrm{in} \hd_{\om_\mathrm{in}}^{\dag}-\sqrt{2\pi}E_\mathrm{in}^* \hd_{\om_\mathrm{in}}
\right)|vac\ra,
\label{eq:inivec}
\eea
where $|vac\ra$ is the overall vacuum state. 
Note that this is an eigenstate of the noise operators: 
$\hGam^{(3e)}(t)|\psi_i\ra = \sqrt{2\pi}\xi_{\om_\mathrm{in}}^{(3e)}\cE_\mathrm{in}(t)|\psi_i\ra$
and $\hGam^{(j)}(t)|\psi_i\ra=0$ for $j=1e, 1i, 3i$, and $q$.

\subsection{Density matrix elements} 
The Heisenberg equation for a system operator is given 
by Eq.~(\ref{eq:dSdt2}) with the dissipators and the noise operators
corresponding to the five decay channels ($1e$, $1i$, $3e$, $3i$ and $q$). 
The equation of motion for $s_{ij}(t)=\la\psi_i|\hs_{ij}(t)|\psi_i\ra$, 
which is identical to the density matrix element 
$\rho_{ji}(t)=\la j|\hat{\rho}(t)|i\ra$ in the Schr\"odinger picture, 
is then given by
\bea
\frac{d}{dt} s_{ij}
&=&
\sum_{m,n}\eta^{(1)}_{ijmn} s_{mn}
+\cE_\mathrm{in}^*(t) \sum_{m,n}\eta^{(2)}_{ijmn} s_{mn}
+\cE_\mathrm{in}(t) \sum_{m,n}\eta^{(3)}_{ijmn} s_{mn}, 
\label{eq:sij2}
\eea
where the coefficients $\eta^{(1,2,3)}_{ijmn}$ are given by
\bea
\eta^{(1)}_{ijmn} 
&=&
i(\ve_i-\ve_j)\delta_{im}\delta_{jn} + 
x_{mi}x_{jn}(h_{nj}^{(1)}+h_{mi}^{(1)*})
-\delta_{im} \left( \sum_l x_{jl}x_{ln}h_{nl}^{(1)} \right)
-\delta_{jn} \left( \sum_l x_{il}x_{lm}h_{ml}^{(1)*} \right)
\nonumber \\
&+& 
y_{mi}y_{jn}(h_{nj}^{(3)}+h_{mi}^{(3)*})
-\delta_{im} \left( \sum_l y_{jl}y_{ln}h_{nl}^{(3)} \right)
-\delta_{jn} \left( \sum_l y_{il}y_{lm}h_{ml}^{(3)*} \right)
\nonumber \\
&+& 
z_{mi}z_{jn}(h_{nj}^{(q)}+h_{mi}^{(q)*})
-\delta_{im} \left( \sum_l z_{jl}z_{ln}h_{nl}^{(q)} \right)
-\delta_{jn} \left( \sum_l z_{il}z_{lm}h_{ml}^{(q)*} \right),
\\
\eta^{(2)}_{ijmn} 
&=&
i\sqrt{2\pi}\xi_{\om_d}^{(3e)}(y_{mi}\delta_{jn}-y_{jn}\delta_{im}),
\\
\eta^{(3)}_{ijmn} &=& (\eta^{(2)}_{ijmn})^*,
\eea
where $h_{ij}^{(1)}=h_{ij}^{(1e)}+h_{ij}^{(1i)}$ and 
$h_{ij}^{(3)}=h_{ij}^{(3e)}+h_{ij}^{(3i)}$.
%
%
In this study, we apply a continuous field and 
observe the stationary response of the system. 
Therefore, we numerically determine 
the stationary solution of these simultaneous equations
by perturbation with respect to $\cE_\mathrm{in}(t)$.
Further details on analysis are presented in Appendix~\ref{app:ss21}. 

\subsection{Photon flux} 
Since we treat a stationary input/output field, 
we quantify the amount of photons by the photon flux, namely,
the rate of incoming/outgoing photons per unit time. 
The input photon flux is evaluated by 
$F_\mathrm{in}=\la \psi_i|\hd_\mathrm{in}^{\dag}(t)\hd_\mathrm{in}(t)|\psi_i\ra$. 
From Eq.~(\ref{eq:inivec}), this quantity reduces to
\bea
F_\mathrm{in} &=& |\cE_\mathrm{in}(t)|^2 = |E_\mathrm{in}|^2.
\eea
In the output port, 
the fluxes of down-converted and unconverted photons
are respectively evaluated by
$F_\mathrm{out}^1=\la \psi_i|\hc_\mathrm{out}^{\dag}(t)\hc_\mathrm{out}(t)|\psi_i\ra$ and  
$F_\mathrm{out}^3=\la \psi_i|\hd_\mathrm{out}^{\dag}(t)\hd_\mathrm{out}(t)|\psi_i\ra$. 
From Eq.~(\ref{eq:cout}) and its counterpart for $\hd_\mathrm{out}$, 
these quantities are given by
\bea
F_\mathrm{out}^1 
&=&
2\pi\sum_{i,j}
\left( \sum_m x_{mi} x_{mj} \xi^{(1e)}_{\ve_{im}} \xi^{(1e)}_{\ve_{jm}} \right)
s_{ij}(t), 
\\
F_\mathrm{out}^3 
&=&
|\cE_\mathrm{in}(t)|^2 + 
2\pi\sum_{i,j}
\left( \sum_m y_{mi} y_{mj} \xi^{(3e)}_{\ve_{im}} \xi^{(3e)}_{\ve_{jm}} \right)
s_{ij}(t) 
+i\sqrt{2\pi}\sum_{i,j}\xi_{\ve_{ji}}^{(3e)} y_{ij} 
\left[ s_{ji}(t) \cE_\mathrm{in}(t)-\mathrm{c.c.}\right].
\eea

\section{numerical results}
\label{sec:6}
In this section, we present the numerical results on the optical response,  
fixing the bare cavity frequencies at $3\om_1=\om_3=2\pi\times 9$~GHz.
For reduction of parameters, we restrict ourselves to the case of 
$g_1=g_3$, $\kap_{1e}=\kap_{3e}$ and $\kap_{1i}=\kap_{3i}$
and denote them by $g$, $\kap_e$ and $\kap_i$, respectively.
Furthermore, regarding the coupling for the $1e$ decay channel for example, 
we assume the following form,
\bea
\xi_k^{(1e)} &=& \theta(k)\theta(k_x-k)\sqrt{\kap_{1e}/2\pi},
\label{eq:xik1e}
\eea
where 
$k_x$ is the cutoff wavenumber. 
Note that this coupling satisfies Eq.~(\ref{eq:kapje}).
We fix $k_x$ at $2\pi\times 20$~GHz
and confirmed that numerical results are mostly insensitive to $k_x$.
The other system-environment couplings are defined similarly
and with the same cutoff wavenumber.
From Eq.~(\ref{eq:hji}), $h_{ji}^{(1e)}$ is analytically given by
\bea
h_{ji}^{(1e)} &=& \frac{\kap_{1e}}{2}\theta(\ve_{ji})\theta(k_x-\ve_{ji})
-\frac{i\kap_{1e}}{2\pi}\log\left(\frac{|k_x-\ve_{ji}|}{|\ve_{ji}|}\right).
\eea
Regarding the input field power, 
we assume the weak-field limit in Secs.~\ref{ssec:opt_kape}--\ref{ssec:intloss},
and discuss the input power dependence in Sec.~\ref{ssec:sat}.

\subsection{Optimal condition for $\kap_e$}\label{ssec:opt_kape}
\begin{figure}
\includegraphics[width=140mm]{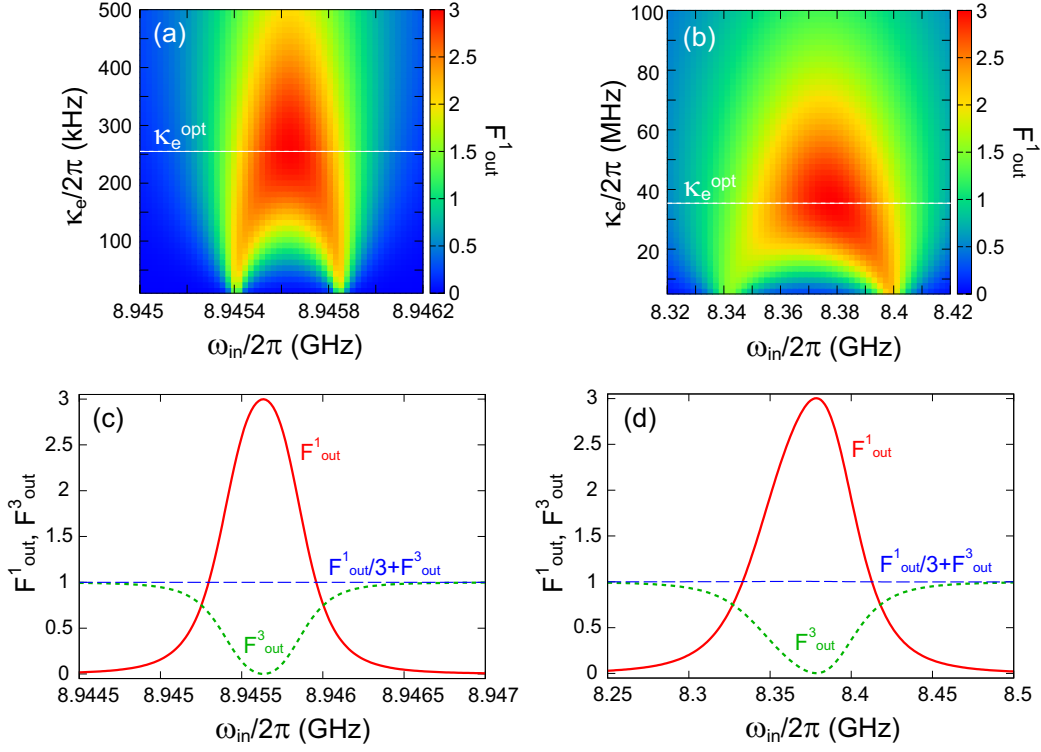}
\caption{
Optimization of $\kap_e$ assuming no intrinsic losses ($\kap_i=\gam=0$). 
The output photon flux is normalized by the input one.
(a)~Dependence of the down-converted flux $F^1_\mathrm{out}$ 
on $\kap_e$ and $\om_\mathrm{in}$, 
for $(g,\om_q)=2\pi\times (0.3, 10.72)$~GHz. 
The optimal point is  
$(\kap_e^\mathrm{opt}, \om_\mathrm{in}^\mathrm{opt})
=2\pi\times (255.0~\mathrm{kHz}, 8.9456~\mathrm{GHz})$. 
(b)~Same plot as (a) for $(g,\om_q)=2\pi\times(1.0, 9.735)$~GHz.
The optimal point is  
$(\kap_e^\mathrm{opt}, \om_\mathrm{in}^\mathrm{opt})
=2\pi\times (35.4~\mathrm{MHz}, 8.378~\mathrm{GHz})$.
(c)~Cross section of (a) at $\kap^\mathrm{opt}_e$: 
down-converted flux $F^1_\mathrm{out}$ (solid), 
unconverted flux $F^3_\mathrm{out}$ (dotted), 
and $F^1_\mathrm{out}/3+F^3_\mathrm{out}$ (thin dashed).
(d)~Cross section of (b) at $\kap^\mathrm{opt}_e$.
}
\label{fig:kapopt}
\end{figure}

First, we search the optimal value of $\kap_e$ 
assuming no intrinsic losses ($\kap_i=\gam=0$).
Figures~\ref{fig:kapopt}(a) and (b) show the dependence of 
the down-converted flux $F^1_\mathrm{out}$ on $\kap_e$
and $\om_\mathrm{in}$, fixing $g$ and $\om_q$. 
It is observed that $F^1_\mathrm{out}$ has two peaks for small $\kap_e$. 
This is due to the Rabi splitting of $|g30\ra$ and $|g01\ra$, 
and the frequency difference of the two peaks 
agrees with $2\geff$ in Fig.~\ref{fig:anti}(c).
$F^1_\mathrm{out}$ is maximized for a larger $\kap_e$, 
at which the two peaks become spectrally indistinguishable. 
Therefore, the optimal condition for the external loss rate of the cavity
is given by $\kap^\mathrm{opt}_e \sim \geff$. 
We confirm in Appendix~\ref{app:impmatch} that 
this condition is identical to the impedance-matching condition
of a linear optical system composed of oscillators and waveguides.
Actually, we can confirm in Figs.~\ref{fig:kapopt}(a) and (b) that
$\kap^\mathrm{opt}_e$ is $2\pi\times 255$~kHz (35.4~MHz) 
for $g=2\pi\times$0.3~GHz (1.0~GHz).  
This is almost identical to $\geff=2\pi\times$223~kHz (29.2~MHz) in Fig.~\ref{fig:anti}(c).

Figures~\ref{fig:kapopt}(c) and (d) are the cross section of 
Figs.~\ref{fig:kapopt}(a) and (b) at $\kap^\mathrm{opt}_e$. 
It is observed that the deterministic down-conversion
($F^1_\mathrm{out} \approx 3F_\mathrm{in}$ and $F^3_\mathrm{out} \approx 0$) 
is attained regardless of the value of $g$, 
when the input photon frequency $\om_\mathrm{in}$ is optimally chosen. 
Furthermore, reflecting the absence of intrinsic loss channels, 
we can also confirm the energy conservation, 
$F^1_\mathrm{out}/3+F^3_\mathrm{out} \approx F_\mathrm{in}$, 
for any input photon frequency.
However, by carefully examining the numerical results,  
this conservation law is slightly broken at the order 
of $10^{-5}$ [$10^{-3}$] in Fig.~\ref{fig:kapopt}(c) [Fig.~\ref{fig:kapopt}(d)]. 
We attribute the main reason for this slight discrepancy 
to the free-evolution approximation [Eq.~(\ref{eq:fev})], 
whose validity is gradually lost for larger dissipation rates. 

\subsection{Qubit detuning}\label{ssec:unopt_qubit}
\begin{figure}[h]
\includegraphics[width=130mm]{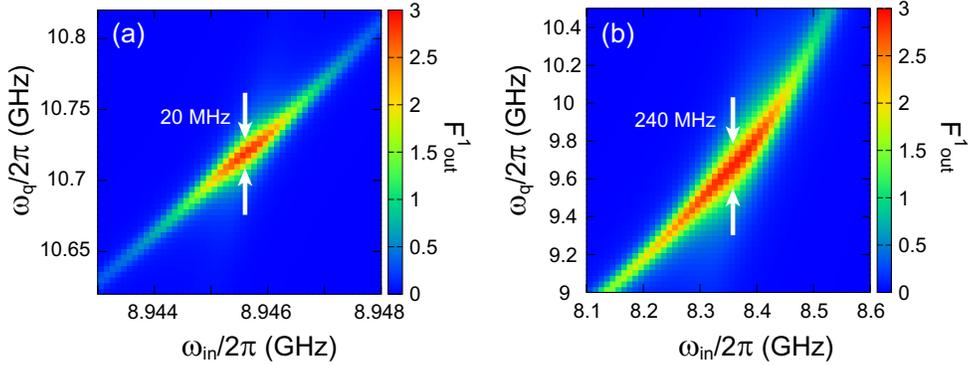}
\caption{
Effect of qubit detuning.
(a)~Dependence of the down-converted flux 
$F^1_\mathrm{out}$ on $\om_q$ and $\om_\mathrm{in}$,
for $g=2\pi\times 0.3$~GHz and $\kap_e=2\pi\times 255$~kHz.
$\kap_i=\gam=0$ is assumed.
(b)~The same plot as (a) for $g=2\pi\times 1.0$~GHz and $\kap_e=2\pi\times 35.4$~MHz.
}
\label{fig:qudet}
\end{figure}
Here, assuming again the absence of intrinsic losses, 
we observe the effects of the qubit detuning from its optimal value.
Figure~\ref{fig:qudet} shows the dependence of 
the down-converted flux $F^1_\mathrm{out}$ on $\om_q$ and $\om_\mathrm{in}$, 
fixing $\kap_e$ at its optimal value 
[255~kHz in (a) and 35.4~MHz in (b)]. 
It is observed that, as the qubit-cavity coupling $g$ increases,
the deterministic down-conversion becomes more robust against the qubit detuning.
This is because of  
the increase of the optimal qubit linewidth for larger $g$. 
The allowed qubit detuning (the full width in $\om_q$ at the half maximum 
of the cross sectional plot at the optimal $\om_\mathrm{in}$) 
is about 20~MHz (240~MHz) for $g=2\pi\times 0.3$~GHz (1.0~GHz).

\subsection{Intrinsic losses}\label{ssec:intloss}
\begin{figure}[h]
\includegraphics[width=130mm]{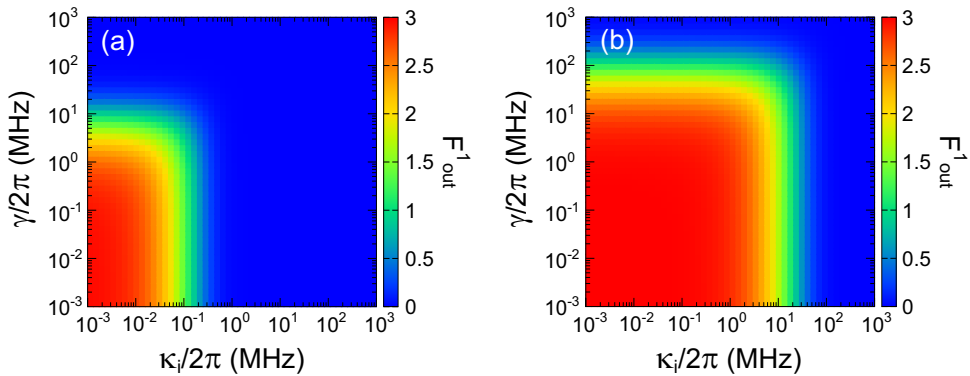}
\caption{
Effect of intrinsic losses. 
(a)~Dependence of the down-converted flux $F^1_\mathrm{out}$ on $\gam$ and $\kap_i$,
for $(g, \om_q, \om_\mathrm{in})=2\pi\times(0.3, 10.72, 8.9456)$~GHz and 
$\kap_e=2\pi\times 255$~kHz.
(b)~The same plot as (a), for 
$(g, \om_q, \om_\mathrm{in})=2\pi\times(1.0, 9.735, 8.378)$~GHz 
and $\kap_e=2\pi\times 35.4$~MHz.
}
\label{fig:loss}
\end{figure}
Here, we investigate the effects of intrinsic losses of the qubit and cavity.
Figure~\ref{fig:loss} shows the dependence of 
the down-converted flux $F^1_\mathrm{out}$ on $\gam$ and $\kap_i$, 
fixing the other parameters at their optimal values. 
When $g=2\pi\times$0.3~GHz, the deterministic down-conversion 
is highly vulnerable to the intrinsic losses. 
The condition for achieving 50\% conversion ($F^1_\mathrm{out}>1.5$) is
$\kap_i\lesssim 2\pi\times$92.9~kHz (intrinsic quality factor 
$Q_i \gtrsim 3.23 \times 10^4$ for the first cavity mode)
and $\gam \lesssim 2\pi\times$5.55~MHz (lifetime $T_1 \gtrsim 28.7$~ns).
These conditions are drastically relaxed for $g=2\pi\times$1.0~GHz:
$\kap_i\lesssim 2\pi\times$12.7~MHz ($Q_i \gtrsim 236$)
and $\gam \lesssim 2\pi\times$70.5~MHz ($T_1 \gtrsim 2.26$~ns).
We observe that the condition for the cavity is tighter than that for the qubit.
This is because, in the present phenomenon, 
the qubit excited state is used only virtually 
to realize the effective coupling between $|g01\ra$ and $|g30\ra$ states.

\subsection{Dependence on input photon rate}\label{ssec:sat}
\begin{figure}[t]
\includegraphics[width=140mm]{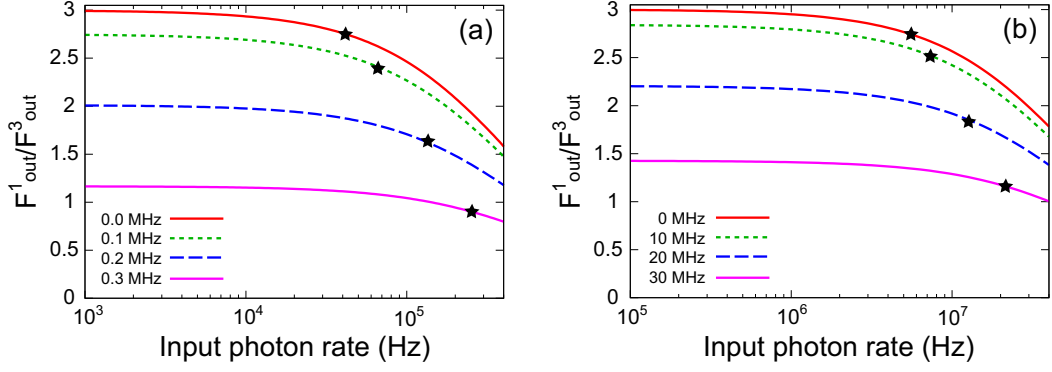}
\caption{
Dependence of the down-conversion efficiency on the input photon rate.
Detuning of the input photon frequency from the optimal one,
$(\om_\mathrm{in}-\om_\mathrm{in}^\mathrm{opt})/2\pi$, is indicated.
Star symbols represent the onset of saturation [Eq.~(\ref{eq:onset})].
(a)~Results under the optimal condition in Fig.~\ref{fig:kapopt}(a). 
(b)~Results under the optimal condition in Fig.~\ref{fig:kapopt}(b). 
}
\label{fig:sat}
\end{figure}

In the previous subsections, we discussed 
the down-conversion efficiency assuming a low input photon rate, 
in other words, the linear-response limit. 
Here, we observe the conversion efficiency for a higher input photon rate. 
In Fig.~\ref{fig:sat}, we plot the dependence of the conversion efficiency 
on the input photon rate for various detuning of the input field. 
We observe that the efficiency decreases gradually for higher input photon rate. 
This is due to saturation of the atom-cavity system, 
which originates from the nonlinearity of the qubit. 
The star symbols in Fig.~\ref{fig:sat} represent the onset of saturation, 
which is given by
\bea
F_\mathrm{in} & \sim & \frac{(\kap_e/2)^2+(\Delta\om)^2}{10\kap_e},
\label{eq:onset}
\eea
where $\Delta\om$ is the detuning of the input field frequency 
from its optimal value. This is derived as follows.
When one applies a monochromatic field $\cE(t)=E_\mathrm{in} e^{-i\om_\mathrm{in} t}$
to an empty one-sided cavity with an external decay rate $\kap_e$, 
the mean intracavity photon number $\overline{n}$ 
is proportional to the drive photon rate $F_\mathrm{in}=|E_\mathrm{in}|^2$ 
and is given by $\overline{n}=\kap_e F_\mathrm{in}/|\kap_e/2+i\Delta\om|^2$.
In the present system, the saturation effect due to nonlinearity 
would appear when the cavity is populated substantially. 
If we set this criterion 
at $\overline{n} \sim 0.1$ for example, 
the onset of saturation is estimated by Eq.~(\ref{eq:onset}).
This explains the fact that
the onset of saturation occurs at a higher input photon rate
for a larger detuning.

\section{Summary}
\label{sec:7}
We theoretically proved the possibility of the deterministic 
three-photon down-conversion of itinerant photons 
using a passive ultrastrong cavity QED system, 
in which an atom is coupled to the fundamental and third-harmonic cavity modes.
For this purpose, we developed an input-output formalism 
applicable to highly dissipative cavity QED systems. 
The conditions for the deterministic conversion are as follows:
(i)~the frequencies of the qubit and the cavity modes are adequately chosen 
so that the two relevant levels ($|g30\ra$ and $|g01\ra$) are coupled effectively,
and (ii)~the cavity loss rates are adequately chosen 
so that they are comparable to the effective coupling. 
Such down-conversion is characteristic to 
the ultrastrong coupling regime of cavity QED,
considering the upper limit of the intrinsic loss rates of the cavity. 

\section*{Acknowledgments}
The author acknowledges fruitful discussions with 
I. Iakoupov, S. Ashhab, and F. Yoshihara.
This work is supported in part by
JST CREST (Grant No. JPMJCR1775) 
and JSPS KAKENHI (Grant No. 19K03684).

\appendix
\section{Validity of $\boldsymbol{f_\mathrm{app}}$}
\label{app:comp}
\begin{figure}[t]
\includegraphics[width=140mm]{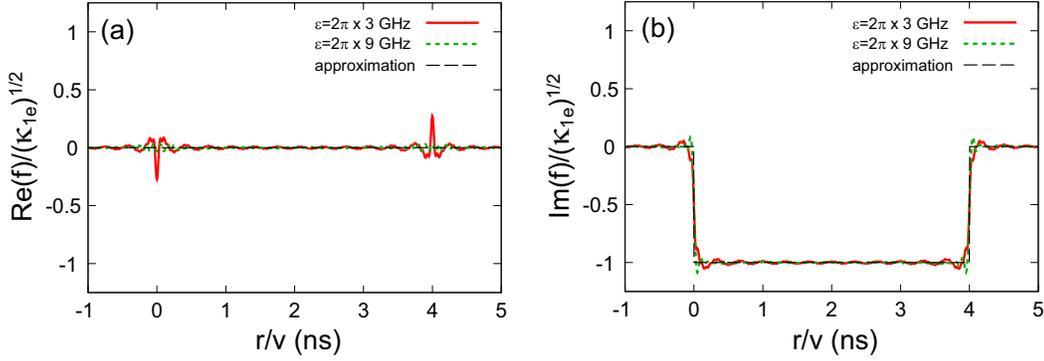}
\caption{
Snapshots of $f$ and $f_\mathrm{app}$ at $t=4$~ns: 
(a)~real and (b)~imaginary parts.
Solid (dotted) lines represent $f$ for $\ve=2\pi\times 3$~GHz (9~GHz), 
and thin dashed lines represent $f_\mathrm{app}$. 
We assume a system-environment coupling of Eq.~(\ref{eq:xik1e}), 
and set the cutoff wavenumber at $k_x=2\pi\times 20$~GHz.
}
\label{fig:frad}
\end{figure}
Here, we numerically compare $f$ and $f_\mathrm{app}$ 
[Eqs.~(\ref{eq:fr1}) and (\ref{eq:fr2})]
that appear when deriving the input-output relation. 
Their snapshots are shown in Fig.~\ref{fig:frad}, 
assuming a concrete form [Eq.~(\ref{eq:xik1e})] 
of the system-environment coupling. 
We confirm that $f_\mathrm{app}$ well approximates $f$
for both the first- and third-harmonic cavity frequencies.

\section{Stationary solution of Eq.~(\ref{eq:sij2})}
\label{app:ss21}
In this Appendix, we present the method to determine 
the stationary solution of Eq.~(\ref{eq:sij2}) perturbatively. 
As the stationary solution, we employ the following form,
\bea
\la s_{ij}(t) \ra &=& \sum_{p,q=0}^{\infty} 
\bars_{ij}^{(p,q)} 
[\cE_\mathrm{in}^*(t)]^p [\cE_\mathrm{in}(t)]^q, 
\label{eq:sij}
\eea
where $\bars_{ij}^{(p,q)}$ is time independent. 
Substituting Eq.~(\ref{eq:sij}) into Eq.~(\ref{eq:sij2}), we have
\bea
\sum_{m,n}\left(
\eta^{(1)}_{ijmn}-i(p-q)\om_\mathrm{in}\delta_{im}\delta_{jn}
\right)\bars_{mn}^{(p,q)}
&=&
-\sum_{m,n}
\left(
\eta^{(2)}_{ijmn}\bars_{mn}^{(p-1,q)}+
\eta^{(3)}_{ijmn}\bars_{mn}^{(p,q-1)}
\right), 
\eea
with the understanding that $\bars_{ij}^{(p,q)}=0$ 
if $p$ or $q$ is negative. 
This is a matrix equation
which we determines $\bars_{ij}^{(p,q)}$
from the lower-order quantities, 
$\bars_{ij}^{(p-1,q)}$ and $\bars_{ij}^{(p,q-1)}$.
Note that this matrix equation is indeterminate for $p=q$. 
Then, we add the normalization condition of the density matrix, 
\bea
\sum_{j=0}^{\infty} \bars_{jj}^{(p,p)} &=& \delta_{p,0}.
\eea

\section{Impedance matching condition}
\label{app:impmatch}
\begin{figure}[t]
\includegraphics[width=80mm]{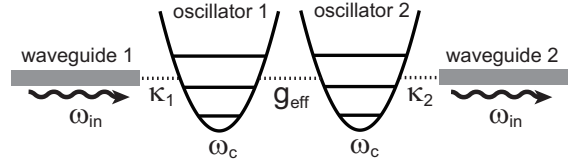}
\caption{
Schematic of the coupled oscillators-waveguides system.
}
\label{fig:cow}
\end{figure}
We consider a linear system composed of 
two harmonic oscillators (oscillator 1 and 2) and 
two waveguides (waveguide 1 and 2), as depicted in Fig.~\ref{fig:cow}.
The two oscillators, which model the levels 
$|g01\ra$ and $|g30\ra$ of the main text,
have the same resonance frequency $\om_c$
and are coupled with a coupling constant $\geff$. 
Oscillator~$j$ ($j=1,2$) is coupled to waveguide~$j$
with an external decay rate $\kap_j$. 
We denote the annihilation operator of oscillator~$j$ by $\ha_j$,
and the input and output field operators of waveguide~$j$ 
by $\hb_{\mathrm{in},j}$ and $\hb_{\mathrm{out},j}$, respectively. 
The Heisenberg equations for the two oscillators
and the input-output relations are given by
\bea
\frac{d}{dt}\ha_1 &=& (-i\om_c-\kap_1/2)\ha_1 
-i\geff\ha_2-i\sqrt{\kap_1}\hb_{\mathrm{in},1},
\\
\frac{d}{dt}\ha_2 &=& (-i\om_c-\kap_2/2)\ha_2 
-i\geff\ha_1-i\sqrt{\kap_2}\hb_{\mathrm{in},2},
\\
\hb_{\mathrm{out},1} &=& \hb_{\mathrm{in},1}-i\sqrt{\kap_1}\ha_1,
\\
\hb_{\mathrm{out},2} &=& \hb_{\mathrm{in},2}-i\sqrt{\kap_2}\ha_2.
\eea
We apply a classical monochromatic field 
at frequency $\om_\mathrm{in}$ and amplitude $E_\mathrm{in}$
through waveguide~1, and apply no field through waveguide~2.
Namely, $\la \hb_{\mathrm{in},1} \ra = E_\mathrm{in} e^{-i\om_\mathrm{in} t}$ and 
$\la \hb_{\mathrm{in},2} \ra = 0$. 
Then, the equations of motion for the cavity and waveguide amplitudes are given by
\bea
\frac{d}{dt}\la\ha_1\ra &=& 
(-i\om_c-\kap_1/2)\la\ha_1\ra -i\geff\la\ha_2\ra
-i\sqrt{\kap_1}\la\hb_{\mathrm{in},1}\ra,
\label{eq:ap1}
\\
\frac{d}{dt}\la\ha_2\ra &=& 
(-i\om_c-\kap_2/2)\la\ha_2\ra -i\geff\la\ha_1\ra, 
\label{eq:ap2}
\\
\la\hb_{\mathrm{out},1}\ra &=& \la\hb_{\mathrm{in},1}\ra-i\sqrt{\kap_1}\la\ha_1\ra,
\\
\la\hb_{\mathrm{out},2}\ra &=& -i\sqrt{\kap_2}\la\ha_2\ra.
\eea
The stationary solution is readily obtained 
by the replacement of $d/dt \to -i\om_\mathrm{in}$ 
in Eqs.~(\ref{eq:ap1}) and (\ref{eq:ap2}).
The transmission coefficient,
$T=\la\hb_{\mathrm{out},2}\ra/\la\hb_{\mathrm{in},1}\ra$, is then given by
\bea
\frac{\la\hb_{\mathrm{out},2}\ra}{\la\hb_{\mathrm{in},1}\ra} 
&=&
\frac{i\sqrt{\kap_1\kap_2} \geff}{\kap_1\kap_2/4+\geff^2}.
\eea
The impedance-matching condition, $|T|=1$, 
reduces to $\sqrt{\kap_1\kap_2}=2\geff$.
This is in agreement with the optimal condition 
of the external cavity decay rate, $\kap_e^\mathrm{opt} \sim \geff$, 
derived in Sec.~\ref{ssec:opt_kape}.




\begin{thebibliography}{99}
\bibitem{fn3}
Since the rigorous real-space representation is given by
$\hc_r=\tc_r+\tc_{-r}$ for $r>0$, 
the commutation relation for $\tc_r$ is given by
$[\tc_r+\tc_{-r}, \tc_{r'}+\tc_{-r'}]=\delta(r-r')$ for $r,r'>0$.
\end{thebibliography}
\end{document}